\documentclass[sigconf]{aamas}

\setcopyright{ifaamas}
\acmConference[AAMAS '22]{Proc.\@ of the 21st International Conference
on Autonomous Agents and Multiagent Systems (AAMAS 2022)}{May 9--13, 2022}
{Online}{P.~Faliszewski, V.~Mascardi, C.~Pelachaud,
M.E.~Taylor (eds.)}
\copyrightyear{2022}
\acmYear{2022}
\acmDOI{}
\acmPrice{}
\acmISBN{}

\newcommand{\citepos}[1]{\citeauthor{#1}'s \citep{#1}}

\usepackage{xcolor}
\definecolor{labelcolor}{cmyk}{0.22,0.10,0.10,0.10}
\definecolor{listbackgroundcolor}{cmyk}{0.10,0.10,0.05,0.05}
\definecolor{listbackgroundcolorlight}{rgb}{0.91,0.92,0.94}
\definecolor{colorEntityBack}{rgb}{0.83,0.83,0.83}
\definecolor{colorPolicyBack}{rgb}{0.91,0.94,0.94}
\definecolor{colorApproachBack}{rgb}{0.93,0.93,0.97}
\usepackage{listings}
\usepackage[inline]{enumitem}

\newenvironment{syn}[1]{\begin{list}{{\footnotesize\sc
	\theenumi.}}{\usecounter{enumi}
      \settowidth{\labelwidth}{{\footnotesize\sc #19}}
      \setlength{\itemsep}{0pt}
      \setlength{\topsep}{0pt}
      \setlength{\parsep}{0pt}
      \setlength{\leftmargin}{\labelwidth}
      \addtolength{\leftmargin}{2.0\labelsep}}}{\end{list}}
\newcounter{syntax} 
\newcommand{\bsyn}{\begin{syn}{L}\setcounter{enumi}{\value{syntax}}\renewcommand{\theenumi}{L$_{\arabic{enumi}}$}}
\newcommand{\esyn}{\setcounter{syntax}{\value{enumi}}\renewcommand{\theenumi}{\arabic{enumi}.}\end{syn}}

\DeclareRobustCommand{\nUmErAL}[1]{}

\newcommand{\fsf}[1]{{\small{{\textsf{#1}}}}}
\newcommand{\fsl}{\textsl}

\newcounter{mpscount}
\newcounter{akccount}
\newcounter{shccount}

\newcommand{\mname}[1]{\fsl{#1}}
\newcommand{\pname}[1]{\fsl{#1}}
\newcommand{\rname}[1]{\textsc{#1}}
\newcommand{\paraname}[1]{\fsf{#1}}

\usepackage{xspace}
\newcommand{\msf}{\mathsf}
\DeclareMathAlphabet{\mathsl}{OT1}{ptm}{m}{sl}
\newcommand{\msl}{\mathsl}
\newcommand{\ulc}{\ulcorner}
\newcommand{\urc}{\urcorner}
\newcommand{\inn}{\ensuremath{\ulc\msf{in}\urc}\@\xspace}
\newcommand{\out}{\ensuremath{\ulc\msf{out}\urc}\@\xspace}
\newcommand{\nil}{\ensuremath{\ulc\msf{nil}\urc}\@\xspace}
\newcommand{\any}{\ensuremath{\ulc\msf{any}\urc}\@\xspace}
\newcommand{\opt}{\ensuremath{\ulc\msf{opt}\urc}\@\xspace}

\newcommand{\lra}{\mbox{$\longrightarrow$}}  

\acmSubmissionID{377}

\title{Pippi: Practical Protocol Instantiation}

\begin{document}
\author{Samuel H.~Christie V}
\affiliation{%
  \institution{North Carolina State University}
  \city{Raleigh}
  \state{NC}
  \country{USA}
}
\email{schrist@ncsu.edu}

\author{Amit K.~Chopra}
\affiliation{
  \institution{Lancaster University}
  \city{Lancaster}
  \country{UK}
}
\email{amit.chopra@lancaster.ac.uk}

\author{Munindar P.~Singh}
\affiliation{
  \institution{North Carolina State University}
  \city{Raleigh}
  \state{NC}
  \country{USA}
}
\email{mpsingh@ncsu.edu}

\begin{abstract}
A protocol specifies interactions between roles, which together constitute a multiagent system (MAS).  Enacting a protocol presupposes that agents are bound to the its roles. Existing protocol-based approaches, however, do not adequately treat the practical aspects of how roles bindings come about.

Pippi addresses this problem of MAS instantiation. It proposes the notion of a metaprotocol, enacting which instantiates a MAS suitable for enacting a given protocol. Pippi demonstrates the subtleties involved in instantiating MAS arising from protocol composition, correlation, and decentralization. To address these subtleties and further support practical application patterns, we introduce an enhanced protocol language, with support for parameter types (including role and protocol typed parameters, for metaprotocols), interface flexibility, and binding constraints. We discuss the realization of our approach through an extended agent architecture, including the novel concept of a MAS adapter for contact management.
We evaluate Pippi's expressiveness by demonstrating common patterns for agent discovery. 

\end{abstract}

\maketitle


\section{Introduction}
Multiagent systems (MASs) can be specified from multiple perspectives, including organizations, commitments, and interactions.
Yet, from a practical standpoint, how do agents come together to instantiate a MAS?

We focus on multiagent systems specified through interaction protocols \citep{TSE-05}, which are characterized by the messages that the agents send to and receive from one another and the constraints on the ordering and occurrence of those messages. For such a specification, the agents take on \emph{roles} in a protocol, which specifies the messages each role may send and receive.
Most protocol specification languages focus on the interaction itself and the order in which messages may be sent and received, with little consideration of how the roles are bound to agents.
From a theoretical standpoint, this is a reasonable assumption; role binding can be left as an implementation detail.
But for practical applications of interaction protocols, it is important to consider how each agent learns of its role and how each agent learns the bindings of the other roles it must interact with.

Composition is a longstanding and challenging theme for practical applications \citep{EIM92}. Protocol composition presents crucial challenges for both modeling and verification \citep{AAMAS-18:atomicity} because protocols focus on interaction and interaction in a decentralized setting is nontrivial to deal with.
Traditional protocol languages rigidly specify the sequences of events that agents can observe, making composition difficult.
Information-based protocol languages such as Splee \citep{AAMAS-17:Splee} and BSPL \citep{AAMAS-BSPL-12} support composition but have their own limitations.
BSPL protocols have fixed conditions for completion, meaning that protocols only have two states, complete and incomplete, and must therefore hide any nuances.
BSPL and Splee protocol key constraints require a composition to have a common key, imposing on their constituent protocols that would otherwise have their own keys.
Finally, current implementations propagate meaning from the outermost composition into each constituent protocol by substituting the parameter names; from a conceptual point of view, this is correct (the meaning of each constituent does change according to its use in the composition), but for modularity each agent should only need to know the protocols it supports, not the compositions they are used in.

We adopt the name Pippi (after the famous character in children's literature) for our approach because of its similarity with PPI (for Practical Protocol Instantiation).

\subsection{Scenario: Wedding}

To illustrate some of these concepts, we adopt a marriage scenario.
The scenario starts with one agent, the proposer, who would like to get married to another.
The proposer is initialized as part of a broader MAS context and can select potential partners from that context.
Once the proposer has selected a partner, they then propose marriage; that is, they propose that the two agents enact a ceremony (protocol) together that will produce a social result of marriage.
When the proposee has agreed to go through the ceremony (possibly selecting which form of ceremony they will enact), they can begin; note that this agreement to enact the ceremony is \emph{not} the same thing as committing to the marriage.

As they enact the marriage ceremony, they will need to bring in other agents to participate; for example, in a court wedding, they need a judge to officiate and a witness to sign the license.
It should be possible to ask someone to witness at the last second, and the witness shouldn't need to know any details about the rest of the ceremony.

\subsection{Objectives and Novelty}

The scenario above presents challenges, which we address:

\subsubsection*{Dynamic Role Binding}
How do agents come together at runtime to enact a protocol?

In the wedding scenario, the proposer starts by inviting their prospective spouse, a judge, and a witness. The proposer must communicate not only each invitee's role in the protocol but also the roles of the other agents they must interact with.
The proposer may prefer to use \emph{late binding} to invite those agents only when their role becomes necessary, to protect privacy for all parties, and avoid overhead in case the enactment is terminated early.

\subsubsection*{Composition}
How can we ensure that exactly the right amount of information is shared between the constituent protocols in a composition?

The wedding scenario can be decomposed into subprotocols, e.g., a specific wedding ceremony and a generic witness protocol.
How do we ensure \emph{independence}, that is, that each agent need only be aware of the protocol it is enacting?
A witness participating in a wedding ceremony need only be aware of the generic witness protocol and should not need to implement support for (or even be aware of) the broader ceremony.

\subsection{Contribution: Our Solution Conceptually}
To negotiate which protocol to enact and which agents to play the roles, agents need the ability to discuss these objects.
We introduce the concept of \emph{metaprotocols}, which are protocols whose parameters refer to the elements of another protocol.

Although metaprotocols can be implemented using normal protocols with special interpretation for some of the parameters, we introduce protocol language extensions to make those meanings explicit.
We borrow role parameters from Splee \citep{AAMAS-17:Splee}, and add support for parameter type declarations.
Parameters with the type \texttt{protocol} may be used as variables for more flexible composition.

\section{The Protocol Language}

BSPL \citep{AAMAS-BSPL-11}, the foundational information protocol language, did not address role binding, assuming that roles were bound prior to protocol instantiation.
Splee \citep{AAMAS-17:Splee} supported treating roles as parameters for dynamic binding but did not address practical instantiation challenges.

We now introduce the core aspects of our protocol language, illustrated through specifications of the scenario.

\subsection{Syntax}
The formal syntax of our language, derived from Splee and BSPL, is given in Table~\ref{syn:Pippi}, and described in detail below.
A superscript of $+$ indicates one or more repetitions, superscript $*$ indicates zero or more, and $\lfloor$ and $\rfloor$ delimit expressions, which are optional when without a superscript.

\begin{table}[t]
\centering
\small
\caption{Pippi Syntax}
\label{syn:Pippi}
\begin{tabular}{@{ }l@{ }l@{ }l@{ }l}\toprule
\ref{syn:Specification} & Spec & $\lra$ & $\lfloor \msl{Name}\text{:} \msl{Value}\ \rfloor^*$
  $\lfloor \msl{Protocol}|\msl{Message}\rfloor^+$\\
\ref{syn:Protocol} & Protocol & \lra &
  $\msl{Name}(\msl{ParamExpr})$\\
  & & & \{
  \ $\lfloor\msf{private}\ \msl{Parameter}^+ \rfloor\ \msl{Reference}^+\ \}$\\
\ref{syn:ParamExpr} & ParamExpr & \lra & $\msl{ParamClause}\ \lfloor , \msl{ParamClause} \rfloor^* (\ \msl{ParamExpr}\ )$ \\
\ref{syn:Parameter} & Parameter & \lra &
  $\lfloor
  \lfloor \msl{Qualifier} \rfloor\ 
  \msl{Name}^+\ 
  \lfloor \text{:}\ \msl{Constraint}^+ \rfloor
  \rfloor^+$ \\
\ref{syn:Reference} & Reference & \lra & $\msl{Name}(\msl{ParamExpr})\ |\ \msl{Message}$ \\
\ref{syn:Message} & Message & \lra &
  $\lfloor\!\msl{Ad}\!\rfloor\ \msl{Name}\!  \mapsto\! \lfloor\!\msl{Ad}\!\rfloor\ \msl{Name}^+\text{:}\
  \msl{Name}[\msl{ParamExpr}]$ \\
\ref{syn:Adornment} & Ad & \lra & $\msf{in}\ |\
  \msf{nil}\ |\ \msf{out} |\ \msf{any}\ |\ \msf{opt}$ \\
\ref{syn:Qualifier} & Qualifier & \lra &
  $\lfloor \msf{key} \rfloor \lfloor \msf{local} \rfloor \lfloor \msf{set} \rfloor$ \\
\ref{syn:Name} & Name & \lra &
  $\lfloor \msl{Prefix}\text{:}\rfloor \msl{(S)tring}$ \\
\ref{syn:Constraint} & Constraints & \lra & 
  $\lfloor \msl{Type} \rfloor
  \lfloor \subseteq\!|\!\in\!|\!=\!|\!<\!|\!> \msl{Name}|\msl{Value} \rfloor$ \\
\ref{syn:Type} & Type & \lra & $\msf{role}\  |\
  \msf{protocol}\ |\ \msl{Name}$
  \\\bottomrule
\end{tabular}
\end{table}

\bsyn
\item\label{syn:Specification} A specification document consists of a preamble and one or more protocols (which may be individual messages).
The preamble contains key/value pairs used to declare common information (such as author or default prefix) or terms for substitution in the document (such as namespaces and IRI abbreviations).

\item\label{syn:Protocol} A protocol declaration consists of a name, a public parameter expression, optional private parameters, and references to constituent protocols or messages.
The public parameters with the $\msf{key}$ qualifier form this declaration's key.
The private parameters can be surrounded by parentheses to spread them across multiple lines.
  
\item\label{syn:ParamExpr} A parameter expression is a conjunction of parameter clauses, written as a (possibly parenthesized) comma-separated list.

\item\label{syn:ParamClause} A parameter clause is a disjunction of parameter expressions or an adorned parameter.
If an adornment is not provided, it should be inferred in a reference or default to \inn.

\item\label{syn:Parameter} A parameter has a name and optional qualifiers and constraints.
Multiple names can be included, separated by commas, as shorthand for repeating the same adornments, qualifiers, and constraints, provided all the names are on the same line or parenthesized.

\item\label{syn:Reference} A reference to a protocol  may consist of a name appended by a parameter expression matching the protocol's declaration.
The name must be the name of a protocol or the name of a parameter whose type is protocol.

\item\label{syn:Message} A message schema consists of a name, sending role and receiving role(s), and a parameter expression.

\item\label{syn:Adornment} An adornment is usually either \inn or \out; \nil indicates an unknown parameter, \any means either \inn or \out, and \opt means the parameter is optional.

\item\label{syn:Qualifier} Qualifiers indicate if a parameter is any combination of key, local, and $\msf{set}$.
The order of the qualifiers is irrelevant.
  
\item\label{syn:Name} Names are strings with an optional namespace prefix separated by a colon and an optional abbreviation in parentheses (the location of the abbreviation in the string does not matter).
Disallowed characters include: [ ()\{\}[]:;,] (whitespace, brackets, and separators)

\item\label{syn:Constraint} Constraints describe the possible values that a parameter binding may have, expressed as a semicolon-separated list.
Example constraints include type, enumerated values, relationships with other parameters, and so on.

\item\label{syn:Type} The type of a parameter may be $\msf{role}$, $\msf{protocol}$, or some name that can be resolved using the context and prefixes declared in the specification.
\esyn

\subsection{Illustration}

\begin{lstlisting}[caption={Generic Witness Protocol},label={bspl:witness}]
Witness(in J, W: role, in key cID, out sig or out objection) {
  private req
  J -> W: RequestApproval[in cID, out req] %\label{line:witness-request}% 
  W -> J: Approve[in cID, in req, out sig]
  W -> J: Object[in cID, in req, out objection]
}
\end{lstlisting}

The \pname{Witness} protocol, given in Listing~\ref{bspl:witness}, shows a simple protocol involving five parameters: \rname{J} (Judge) and \rname{W} (Witness) are roles, \paraname{cID} is the key that uniquely identifies each enactment (which is adorned \inn and therefore bound externally), and \paraname{sig} (signature) and \paraname{objection} are two parameters that may be bound during an enactment.
The public parameters declare the completion requirements for an enactment: when all public parameters are bound, the enactment is complete.
In \pname{Witness}, we demonstrate syntax for complex Boolean completion formulas; only one of \paraname{sig} or \paraname{objection} needs to be bound to complete the enactment.

The body of the protocol lists the subprotocols it is composed of; a message is an elementary protocol.
The \mname{RequestApproval} message on line~\ref{line:witness-request} is sent by \rname{J} (the role before the arrow) to \rname{W}.
Its payload contains two parameters, \paraname{cID} and \paraname{req}; these parameters represent the information conveyed by the message.
The parameter \paraname{cID} is adorned \inn, and so must be bound before \mname{RequestApproval} can be sent.
Conversely, \paraname{req} is \out, and so must \emph{not} already be bound; it is bound when sending \mname{RequestApproval}.
Parameter bindings are functionally dependent on their keys; \paraname{req} may have only one binding for each value of \paraname{cID}.

Messages may be sent when they are enabled by the satisfaction of their causality constraints as specified by their parameter adornments.
Only emissions are constrained; a message may be received any time after it has been sent.

\begin{lstlisting}[caption={Court Wedding},label={list:court-wedding}]
CourtWedding(in Propose(R), Propose(E), (J)udge, (W)itness: role,
  any key cID,
  (out vowR, out vowE, out license) or out objection) {
  private signature
  J -> R, E: Ask[any cID, out questions]
  E -> J, R: EVow[in cID, in questions, out vowE]
  R -> J, E: RVow[in cID, in questions, out vowR]
  Witness(J, W, cID, signature or objection)
  J -> E, R: Marry[in cID, in vowE, in vowR, in signature, out license]
}
\end{lstlisting}

Listing~\ref{list:court-wedding} gives a protocol describing a court wedding.
Note that all of the role parameters are adorned \inn; they must be bound outside the context of the protocol.
The key \paraname{cID} is adorned \any, which means it can be either \inn or \out.
In addition to the features used in \pname{Witness}, \pname{CourtWedding} references \pname{Witness}, and handles the two outcomes separately.
If the witness provides a signature, then the judge can perform the marriage; otherwise, the protocol terminates with an objection.
\pname{CourtWedding} exhibits the role abbreviation syntax, e.g., identifying \rname{R} as the abbreviation for \rname{Proposer}.

\section{Dynamic Role Binding}
All that is required to enact a protocol is a group of agents willing and able to play all its roles.
In most existing work on protocols, the role bindings are assumed; perhaps the MAS was hard-coded or statically configured with the agents' roles and endpoints.

For greater flexibility and to properly model real-life systems, agents need to discover new peers and negotiate protocols with them.
Such dynamic role binding is useful even for simple systems, such as where one seller enacts the same purchase protocol with multiple customers.
Explicit semantics for how the seller is introduced to each customer and how those role bindings relate to the enactment information is necessary for correct implementation.

Furthermore, we would like to support not only explicit role binding but also \emph{late} role binding, where some agents begin to play their roles after the protocol has already been initiated.
For example, in a rideshare setting, riders might want to request a different driver if the first is not coming without having to resubmit their destination.
In addition, it is reasonable to delay the involvement of additional parties for efficiency and privacy reasons.
If a customer's bank is notified every time the customer considers a purchase regardless of their final choice, it unnecessarily exposes their actions and wastes the bank's resources.

To address the above concerns, we introduce two patterns of protocol design: metaprotocols and self-contained protocols.

\subsection{Metaprotocols}
Metaprotocols are protocols that specify communication about the elements of another protocol, such as its structure and role bindings.
Roles in a protocol may be represented by parameters in a metaprotocol without any special semantics.

Late role binding is enabled in a metaprotocol by interleaving its enactment with that of its target protocol.
Then, messages binding specific roles need not be sent until they are relevant in the target protocol, though, of course, they may be sent earlier.

A metaprotocol capturing the proposal phase of the marriage scenario might be implemented as follows:
\begin{lstlisting}[caption={Proposal Metaprotocol},label={prot:proposal}]
Proposal(out r0, r1: role, out key mID, out acceptance or out rejection) {
  private (r2, r3: role,
    cID, R, E, J, W,
    ceremony: protocol)
  out r0 -> out r1: Propose[out mID, out ceremony, out R, out E]
  r1 -> r0: Accept[in mID, in ceremony, out acceptance]
  r0 -> r1: Plan[in mID, in ceremony, in accept, out cID]
  ceremony(R, E, J, W, cID, out vowR, out vowE, out result)
  r0 -> out r2: Schedule[in mID, in cID, in E, out J, out date]
  r0 -> out r3: Invite[in mID, in cID, in J, out W, out invitation]
  r1 -> r0: Reject[in mID, in ceremony, out rejection]
}
\end{lstlisting}

The \pname{Proposal} metaprotocol, given in Listing~\ref{prot:proposal}, sets up the roles necessary for enacting a wedding ceremony such as \pname{CourtWedding}.
First, the roles in this protocol are numbered because they have no meaning in themselves; they are placeholders for the purpose of inviting agents to perform the ceremony.
Thus, r0 sends \mname{Propose} to r1, suggesting that they take on the roles R and E in some ceremony---not necessarily in that order.
The ceremony is specified by binding it to the \paraname{ceremony} parameter.
If r1 accepts, r0 may proceed to invite other agents.
Note that \emph{all} of the roles from \pname{CourtWedding} are bound in this protocol, yet late binding is possible because r0 may delay inviting r3 to perform the W role until it becomes relevant.

Metaprotocols can be used to implement protocol negotiation, as \pname{Proposal} demonstrates by using a parameter to reference a protocol: r0 can propose the ceremony to enact, provided it matches the same interface.
However, not all metaprotocols involve negotiation; a simple metaprotocol could simply communicate the relevant role bindings to each participant.

Metaprotocols cleanly separate discussion about the protocol from discussion of its enactment.
They are fully compatible with BSPL even without our syntax enhancements and may be easily written for existing protocols.
Metaprotocols do not reduce flexibility but enable it.
Without a metaprotocol, role binding is not flexible but merely unspecified, leading to tight coupling and hidden complexity in a multiagent system.

\subsection{Metaprotocol Generation}
A simple algorithm for generating a metaprotocol from an existing protocol is as follows:
\begin{enumerate}
    \item Compute a contact graph among the roles, with directed edges following the message transmissions.
    \item Identify the initiating role, possibly with user input.
    \item Using the contact graph compute the rank of each role (distance of each role from the initiator).
    \item Generate a message from the initiator that binds a metaprotocol ID, and another binding the protocol ID.
    \item Generate messages from each agent to their contacts of higher rank (flowing from the initiator), inviting them to the protocol and naming any mutual contacts they have invited.
\end{enumerate}

The above algorithm generates introduction messages for each role according to the contact graph of the original protocol; in this way, agents are responsible for inviting only agents they contact.
The rank-based invitation approach produces a minimal metaprotocol; that is, the smallest number of single-recipient messages that cover all of the invitations.
As such, the complexity of the metaprotocol is no worse than the original protocol; it has approximately one message per role.

The resulting metaprotocol can be customized to support domain requirements.
For example, the metaprotocol for a purchase protocol can be customized so that the seller chooses the transaction ID, even though the customer initiates the interaction.
Or, in the wedding example, the metaprotocol can specify that the proposer will invite the witness even though the judge is the only agent who contacts the witness during the ceremony.

Using the above algorithm on \pname{CourtWedding} and arbitrarily selecting \rname{R} as the initiator, we can derive the following:
\begin{enumerate}
    \item R sends messages to J and E; J sends to R, E, and W; E sends messages to R; W sends to J
    \item R has rank 0 (as the selected initiator), J and E have rank 1, and W has rank 2
\end{enumerate}

In general, the initiating agent could be identified from the protocol by looking at which messages declare the primary keys of the protocol or depend only on external parameters. However, in this particular example, the initiator for the metaprotocol (R) is not the same as the first agent to communicate in the CourtWedding protocol (J), so it must be explicitly selected as the initiator.

With the above contact graph and rank information, we can derive the comprehensive metaprotocol shown in Listing~\ref{list:generated-metaprotocol}.
The \pname{Generated} protocol in Listing~\ref{list:generated-metaprotocol} is simple because the rank sorting prevents \rname{J} from inviting \rname{E} and vice-versa.
\rname{E} is arbitrarily invited first in, e.g., alphabetical order.
Our example algorithm is far from the only possibility; alternative algorithms could generate a multicast message that simultaneously invites multiple agents or generate a more complex metaprotocol to invite them in either order.

\begin{lstlisting}[caption={Generated Metaprotocol for CourtWedding},label={list:generated-metaprotocol}]
Generated(out r0, r1, r2, r3: role,
  out R, E, J, W: role,
  out key mID, out cID,
  out vowR, vowE, out license or out objection) {
  out r0 -> out r1: InviteE[out mID, out cID, out R: role=r0, out E: role=r1]
  r0 -> out r2: InviteJ[mID, cID, R, E, out J: role=r2]
  r2 -> out r3: inviteW[mID, cID, R, E, out W: role=r3]
}
\end{lstlisting}

We used type and value constraints (e.g., '\texttt{role=r2}', where '\texttt{role}' is a type constraint and '\texttt{=r2}' is a value constraint) for the \out role parameters; both are optional but make the protocol clearer.

\subsection{Self-Contained Protocols}

Although the previous approach to protocol configuration is conceptually simple and compatible with existing BSPL semantics, it can be somewhat impractical because every nontrivial protocol must have a corresponding and cumbersome metaprotocol.
Those metaprotocols can add many messages as overhead solely for communicating contact information.

We propose further extensions to BSPL to support more practical, progressive protocol instantiation at the cost of some complexity.

\begin{lstlisting}[caption={Self-Contained Wedding Protocol},label={list:self-contained-wedding}]
Self-Contained-Wedding(out Propose(R), Propose(E): role 
  out key cID, out rejection or
    (out vowR, out vowE, out license or out objection)) {
  private (J)udge: role, (W)itness: role, signature
 
  out R -> out E: Propose[out cID]
  E -> R: Accept[in cID, out acceptance]
  E -> R: Reject[in cID, out rejection]
  R -> out J: Schedule[in cID, in E, in acceptance, out date]
  J -> R, E: Ask[in cID, out questions]
  E -> J, R: EVow[in cID, in questions, out vowE]
  R -> J, E: RVow[in cID, in questions, out vowR]
  R -> out W: Invite[in mID, in cID, in J, out invitation]
  Witness(J, W, cID, out signature or out objection)
  J -> E, R: Marry[in cID, in vowE, in vowR, in signature, out license]
}
\end{lstlisting}

The \pname{Self-Contained-Wedding} protocol in Listing~\ref{list:self-contained-wedding} does not rely on a separate metaprotocol but instead specifies the process and propagation of role binding internally.

Self-contained protocols are not superior to metaprotocols, but may be an optimization if they can, e.g., communicate role bindings alongside other information in the protocol.

\section{Composition}
Composition is another important aspect of information protocols.
Implicitly, agents can participate in multiple protocols at the same time and must keep them separate.
Protocols can be explicitly composed into a larger protocol specification.
Provided all the agents know the composite protocol, composition adds no complexity to an implementation.
However, the following concerns arise with respect to instantiation:
\begin{enumerate*}
\item enabling interface flexibility, so that protocols may share or hide information as necessary with the broader composition
\item agents should not be required to know all compositions containing the protocols they support
\item enabling composition of protocols without needing a common key.
\end{enumerate*}

\subsection{Interface Flexibility}
Existing information protocol languages BSPL and Splee have rigid protocol interfaces; all public parameters must be bound for an enactment to complete.
Hence, protocol designers must encode multiple termination states---such as accepting or rejecting a marriage proposal---in a single result parameter.

\begin{lstlisting}[caption={Information Hiding Example},label={list:info-hiding}]
OpaqueOffer(in B,S: role, in key oID, out item, amount, result) {
  private acceptance, rejection
  S -> B: Offer[oID, out item, out amount]
  B -> S: Accept[oID, item, amount, out result]
  B -> S: Reject[oID, item, amount, out result]
}

Purchase(out B, S: role, out key pID, out transaction) {
  private request, item, amount
  out B -> out S: RFQ[out pID, out request]
  OpaqueOffer(B, S, pID, item, result)
  S -> B: Ship[pID, result, out transaction]
  // no way to distinguish outcomes in composition
  S -> B: Cancel[pID, result, out transaction]
}
\end{lstlisting}

\pname{OpaqueOffer} in Listing~\ref{list:info-hiding} has only a single \paraname{result} parameter in its public parameter line, hiding the actual response of \rname{B}.
If used in a composition such as \pname{Purchase} in the same listing, there is no way to distinguish the outcomes except through internal agent logic applied to the parameter bindings.

However, when protocols are used in a composition the limitation becomes clearer; what if some subsequent actions should be available only after acceptance, and others only after rejection?
For example, the wedding ceremony should be enacted only if the proposee accepts the proposal.
In this case, the ceremony has an explicit dependency on the outcome of the proposal, so it should not be hidden.

Our proposed solution is to enhance protocol interfaces to support more complex Boolean expressions over the protocol's parameters.
Where before the interface was a conjunction over the public parameter bindings, now parameters can be grouped into disjoint clauses.

\begin{lstlisting}[caption={Flexible Interface},label={list:flexible-interface}]
TransparentOffer(in B, S: role, in key oID, out item, amount, out acceptance or rejection) {
  private acceptance, rejection
  S -> B: Offer[oID, out item, out amount]
  B -> S: Accept[oID, item, amount, out acceptance]
  B -> S: Reject[oID, item, amount, out rejection]
}

Purchase(out B,S:role,out key pID,out package or out canceled) {
  private request, item, amount, acceptance, rejection
  out B -> out S: RFQ[out pID, out request]
  TransparentOffer(B, S, pID, item, acceptance or rejection)
  S -> B: Ship[pID, acceptance, out package]
  S -> B: Cancel[pID, rejection, out canceled]
}
\end{lstlisting}

\pname{TransparentOffer} in Listing~\ref{list:flexible-interface} shows how Boolean parameter expressions enable more transparent interfaces that expose multiple termination states, which can then be handled separately in a composition.
The parameters \paraname{acceptance} and \paraname{rejection} are normal parameters joined into a disjunctive clause by the \texttt{or} keyword; only one of the two bindings is necessary for completion.
Thus, an enactment of \pname{TransparentOffer} is complete when the parameters \paraname{oID, item, amount}, and either \paraname{acceptance} or  \paraname{rejection} are bound.

Note that because messages are elementary protocols, they may use Boolean parameter expressions.
However, this flexibility can obscure the meaning of the messages and so should be used with care.
For example, in a wedding or purchase protocol, sending an explicit \mname{Accept} message forms a commitment unambiguously and can be interpreted at the event level without inspection.
By contrast, interpreting a more abstract \texttt{Decide[ID, out acceptance or out rejection]} requires examining the payload.

\subsection{Constituent Protocol Independence}
Semantically speaking, when a protocol is referenced in a composition, its parameters are substituted by the parameters passed in the reference.
Thus, a bank transfer protocol may involve a parameter named \paraname{amount} that is used to transfer \paraname{payment} in a purchase protocol or \paraname{refund} in a protocol for processing returns.
Proper encapsulation and modularity require that the bank is able to enact its role in the transfer without being aware of the broader purpose.
This requirement is violated by current implementations because the bank would need to interpret the substituted parameter names instead of the expected \paraname{amount}.

\begin{lstlisting}[caption={Dependent Representation},label={list:dependent-repr}]
// (D)ebtor, (C)reditor, (B)ank
// original: D -> B: Transfer[in ID, out amount, C]%\label{line:transfer-schema}%
{"ID": <uuid>, "amount": 100, "C": "Creditor"}
// (B)uyer, (S)eller, Ban(K)
//composed: B -> K: Transfer[in pID, out payment, S]%\label{line:transfer-composed}%
{"pID": <uuid>, "payment": 50, "S": "Seller"}
\end{lstlisting}

Listing~\ref{list:dependent-repr} shows two versions of a message schema.
The first schema (line \ref{line:transfer-schema}) is the generic transfer expected by the bank with the parameter \paraname{amount} specifying the amount to be transferred.
Its representation in JSON is given on the following line, with the expected parameter names as dictionary keys.
Using the transfer protocol in a purchase composition would result in substituting the parameters, as shown on line \ref{line:transfer-composed}.
The subsequent JSON representation uses the parameter names from the composition, which may not be understood by the bank.

To better support protocol independence, we propose a simplification of message representations, where each message is encoded as a (name, payload) pair, and the payload is simply an array of unnamed parameters.
The name must uniquely identify the message in the MAS; it could be an IRI or a more compact encoding defined by the MAS for efficiency.
The payload is interpreted by referring to the message schema.

Replacing named parameters with a uniquely identified message schema minimizes the context sensitivity of the message and simplifies interpretation.
That is, referring to the protocol that specifies the message provides the information necessary for decoding the message: which parameters are included in which order.
However, all semantic information is left out of the encoding, so each agent can interpret the message according to its understanding of the composition.
This is similar to message encoding libraries such as Protocol Buffers \citep{Protobufs}, which use an external schema to parse the messages, enabling compact encoding and backward compatibility across schema versions (since names can change without changing the encoding).

\begin{lstlisting}[caption={Independent Representation},label={list:independent-repr}]
// original: D -> B: Transfer[in ID, out amount, C]
[<uuid>,100,"Creditor"]
// composed: B -> K: Transfer[in pID, out payment, S]
[<uuid>,50,"Seller"]
\end{lstlisting}

In Listing~\ref{list:independent-repr}, leaving out the parameter names does not make the schema ambiguous but does make it independent of its use in a composition.
The bank receives the information necessary to execute a transfer without being confused by the buyer's intent to use the transfer as payment for a purchase.

\subsection{Global and Local Keys}
We have added syntax for distinguishing global and local keys.
The purpose of the distinction is to enable \emph{invertible} relationships between protocol keys, without which protocols in composition must share a common key to share information.

\begin{lstlisting}[caption={Noninvertible Key Composition},label={list:noninvertible-keys}]
Noninvertible(B, S: role, key ID, oID, dID,
  item, price, package) {
  out B -> out S: Start[out ID]
  Order(B, S, ID, oID, item, price)
  Deliver(B, S, ID, dID, item, package)
}
\end{lstlisting}
As shown in Listing~\ref{list:noninvertible-keys}, the \pname{Order} and \pname{Deliver} protocols must share an \pname{ID} to correlate the information, which intrusively imposes awareness of the composition on those protocols.
If written independently, each would need only its own key.

An invertible relationship exists between keys if they are uniquely associated with each other; that is, an agent receiving a message containing one key can correlate it to an enactment of another protocol using the other key.
If a relational data model is used, the agent's adapter can construct queries for enactment information using an association table to join information from the two protocols.
Or, the agent's adapter can use a dictionary to map the unique keys to a single shared enactment object.

Listing~\ref{list:invertible-keys} shows a composition where the keys are universally unique.
Because both \paraname{oID} and \paraname{dID} are unique, the \mname{TrackingInfo} message declares an invertible relationship between them, enabling both agents to correlate the messages inside \pname{Deliver} to the broader enactment without needing to propagate a common key.

\begin{lstlisting}[caption={Invertible Key Composition},label={list:invertible-keys}]
Invertible(B, S: role,
  key oID, dID // new default of global keys
  item, price, package) {
  Order(B, S, oID, item, price)
  S -> B: TrackingInfo[in oID, out dID]
  Deliver(B, S, dID, item, package)
}
\end{lstlisting}

Global keys are the default. They are identified by the sole qualifier \texttt{key}, and must be guaranteed unique within the MAS.
We include the \texttt{local} keyword to cover all other kinds of keys (e.g., sequence IDs or timestamps). Such keys are required to be unique only within the context of the global keys.
Thus, a local key must be used in the context of a global key, possibly scoped further by other \inn local keys.
If multiple local keys are \out in the same message, they form a hierarchy based on the order in which they are given.

Listing \ref{list:local-keys} gives an example of how local keys might be used in a protocol.
It specifies a support request protocol, where \rname{B} opens a ticket, specifying a topic and generating a new unique ID.
Then, \rname{B} and \rname{S} can continue replying with an indefinite number of messages, each uniquely identified with reply IDs (possibly sequential or timestamped), bound to \paraname{brID} and \paraname{srID}, respectively.
These response IDs need only be unique within the context of the support request because they are local keys.

\begin{lstlisting}[caption={Local Key Example},label={list:local-keys}]
Support (B, S: role,
  key ID, // global key
  local key brID, srID, // local keys
  topic, closed) {
  B -> S: OpenTicket[out ID, out topic]
  S -> B: SReply[in ID, in topic, out srID, out content]
  B -> S: BReply[in ID, in topic, out brID, out content]
  S -> B: CloseTicket[in ID, out closed]
}
\end{lstlisting}

\section{Realization}

Programming agents to support dynamic role binding and multiple parallel interactions need not be complicated but is more involved than supporting a single protocol.

In previous programming models for information protocols (PoT \citep{Computer-20:PoT}, Bungie \citep{Computer-21:Bungie}, Deserv \citep{ICWS-21:Deserv}), agents adopt a single role in a single protocol.
Architecturally, each agent was instantiated using a protocol adapter configured with the role and protocol that the agent supported.
The adapter was configured with static role bindings that were bound outside of the protocol enactment. However, it may be possible to change these bindings; the adapter supported only a single binding for each role and protocol.
Supporting additional protocols would require the agent to run multiple adapters with separate endpoints.

\subsection{Protocol Adapter}
The protocol adapter must support multiple roles across several protocols and metaprotocols.
Each protocol is loaded in the context of a specification, which may define terms used in the protocols.

The adapter would support a collection of initial protocols (or metaprotocols), along with the roles it is willing to play in those protocols and handlers for those messages.
These initial protocols could be implemented according to patterns such as peer-to-peer introductions or a central registry, among others, which we examine further in Section~\ref{sec:Evaluation}.

During the enactment of the initial protocols, the agent will either propose enactments of protocols as specific roles or receive invitations to take on a role.
Fortunately, the agent need not specify special-case handlers for these invitations any different from other messages; the information structure of the protocols is enough to guide further enactment.

Roles have become parameters bound within the context of a protocol or metaprotocol instance key.
No longer are roles implicitly part of the message schema. A role must either be explicitly bound in sending a message to a new role or provided by the enactment history.
Provided the protocol (or metaprotocol) is verified to be live and enacted without violation or error, the agents will receive the contacts they need to complete their role.
How an agent selects or approves the specific role bindings is left to their internal logic.

\subsection{MAS Adapter}
To better support agent discovery and selection, we propose a new component, the \emph{MAS Adapter}.
The MAS adapter is optional; simple agents (such as web services) may handle each enactment independently.
Provided each role binding is used only within the context of an enactment, the information management provided by the protocol adapter is sufficient.
The MAS adapter is responsible for remembering the contacts and the history of interactions with them beyond a single enactment.

The MAS adapter implements a registry of known agents.
New entries are automatically added as agents are introduced.
Each entry contains information about the protocols and roles that the agent is believed to support, based on the introduction.
Each entry contains information about the roles that the agent has played and the corresponding enactment histories.
Additional information can be added according to domain requirements, such as quality of service ratings.
An agent can query its MAS adapter to find agents to enact protocols with.

\section{Evaluation}
\label{sec:Evaluation}

We now show how the above concepts can be applied toward capturing canonical role-binding patterns in practical decentralized applications.

\subsection{Preconfigured Contacts}
In peer-to-peer applications, it is common for the peers to be preconfigured with the knowledge of some bootstrap nodes to get the application going.
Network applications are typically preconfigured with a DNS server for purposes of resolving domain names.

In our context, preconfigured contacts are those an agent is given before it is initiated. Preconfigured contacts are implemented in the agent internals and used to bind the initial \out roles of the first protocols the agent enacts. This is a simple pattern, but it is a necessary component of all more complex patterns.

\begin{lstlisting}[caption={Preconfigured Contacts Example (JSON)},label={list:preconfigured-contacts}]
{"Seller":["http://storeA.com/agent","http://storeB.com/agent"],
 "Bank": ["http://bank.com/agent"]}
\end{lstlisting}

Listing~\ref{list:preconfigured-contacts} shows how contacts might be declared in a JSON file.
Preconfigured contacts are not hard coded bindings; they are candidates that are bound to roles dynamically according to the protocol structure and agent logic.

\subsection{Central Registry}
The next level of complexity from a collection of preconfigured contacts is a central registry provided as a service by another agent.
Discovering agents from a central registry is common in services.
The Uber application, e.g., uses Uber's (the organization's) central registry of potential drivers to bind as the pickup driver in a particular transaction with a customer. 

A central registry is simple in that it means that an agent configuration requires only a single registry connection. In addition, a registry enables more dynamic and scalable peer discovery than preconfiguration.
As each agent comes online, they can enact a registration protocol, notifying the registry of their existence and willingness to perform specific roles.
To enact a protocol, an agent can first query the registry to discover potential peers.
As the example in Listing~\ref{list:central-registry} shows, the pattern could be implemented as two protocols, where a single well-known agent uses its MAS adapter to remember and recommend contacts to other agents.

\begin{lstlisting}[caption={Central Registry},label={list:central-registry}]
Registration(out A, R: role,
  out key ID, out endpoint, confirmation
  out set protocols: protocol) {
  out A -> out R: Register[out ID, out endpoint, out protocols]
  R -> A: Confirm[in ID, in endpoint, in protocols, out confirmation]
}

Discovery(out Q, R: role,
  out key ID, out protocol: protocol, out set agents: role) {
  out Q -> out R: Query[out ID, out protocol]
  R -> Q: Introduce[in ID, in protocol, out agents]
}
\end{lstlisting}

\subsection{Peer Sharing}
Peer sharing is characterized by the absence of any distinguished system nodes that support discovery.
Peer-to-peer discovery and binding are common in MAS and distributed systems.
For example, it is used in referral networks \cite{TSMC-05}, in the Contract Net \cite{smith:contract-net:1980}), and in leader election protocols.
Further, mesh networks and IoT-based systems typically invoke the peer-to-peer pattern. 

To find a desired peer, each agent checks its own MAS adapter and queries its neighbors.
The exact nature of the peer selection and query process is application-specific, but generally, each query will return more peers; either the desired peer will be among them, or they can be queried in turn.

\begin{lstlisting}[caption={Peer Discovery},label={list:peer-sharing}]
Discover(out P1, P2: role, out key ID,
  out set protocols: protocol, out set neighbors: role) {
  out P1 -> out P2: Query[out ID, out protocols]
  P1 -> P2: Introduce[in ID, in protocol, out neighbors]
}
\end{lstlisting}

In the peer discovery protocol in Listing~\ref{list:peer-sharing}, there is no need for registration; each agent simply needs an initial bootstrap peer to connect to.
Each time an agent asks a peer for neighbors, it naturally introduces itself and reveals the protocol(s) it is interested in.

\section{Discussion}

We conceptualized the problem of instantiating a MAS in terms of binding the roles of the protocol that models the MAS.
Although multiagent systems are typically thought of as being open in the sense of agents dynamically joining and leaving the system, MAS approaches have not paid sufficient attention to the operational aspects of the problem.
Pippi demonstrates how MAS can be instantiated dynamically by showing how a protocol's roles are bound.
Our approach supports dynamic (not hard-coded) and late (just-in-time) role bindings: not all roles in a protocol need to be bound before enacting the protocol.
We discussed a possible realization of our approach and demonstrated how commonly used patterns can be captured in our approach.

\citet{Dastani+03:role-assignment} consider compatibility between agent and role specifications as a basis for an agent's decision whether or not to play a role.
Such internal decision making, although practically relevant, is outside the scope of Pippi. Pippi is concerned with public aspects of decision making as reflected in protocol enactments.
HAPN \cite{Winikoff+18:HAPN} supports dynamic role binding; e.g., an auction's winner is dynamically bound.
It does not give a general metaprotocol-based approach that enables discovery and bindings of roles.
\citet{Grenna-2008:organizations} extended JADE to enable agents to enact roles, but within the context of an organization rather than a protocol.


\citet{Gunay+15:generating} propose a metaprotocol by which agents can reach agreement on the commitment protocol: All agents should accept that they will create the commitments involved.
They assume role bindings (e.g., that there are customer and merchant agents).
Thus, Pippi complements their work.
\citet{McGinnis+Robertson-05:protocol} propose protocols as first-class abstractions for composing open MAS. Pippi agrees with their intuition and demonstrates how role bindings generated as information in the messages of one protocol (the metaprotocol) can be used as roles (the senders and receivers of messages) in another protocol via composition.

Multiagent systems are conceived of as open systems in the sense that agents can join and leave the MAS.
\citet{Mazouzi+02} give an early example of an agent wanting to join several groups and show that abstract specifications can be refined depending on requirements.
Role binding, as we formalize here, is akin to ``joining'' a MAS.
However, because parameter bindings and hence role bindings are immutable, ``leaving'' a MAS is not unbinding the role.
Leaving would be captured by communicating that the agent has left the MAS.
Immutability of information gives the immutability of events, which is necessary for realism.
For example, once bound as a partner in some wedding, an agent is always a partner for that wedding.
Divorce is not an unbinding of the agent to the partner role but a change in the normative relationships \citep{WWW-16:IOSE} between the partners, perhaps via the creation of a new MAS.
For example, two divorced people may have joint custody of their children and may thus function together in a MAS, albeit not the same MAS as when they were married.
Such ideas merit further study.

\citet{Minsky+Murata-03:open-MAS} discuss the robustness and manageability of a MAS modeled as a law-based society \citep{Minsky91}. In considering robustness, \citeauthor{Minsky+Murata-03:open-MAS} bring up both static and dynamic role binding in the law, which is realized as a set of Prolog-like rules. The static bindings are given as facts, and the dynamic bindings are established via events referred to in the rules. Our approach is compatible with rule-based specifications (e.g., commitments or other norms \citep{WWW-16:IOSE}) but captures the general operational aspects of role binding in a decentralized system via protocols.

\citet{Ferrando+19:enactability} consider the enactability \citep{AAAI-Enact-08} of protocols specified in a language for specifying execution traces, where the constraints specify message ordering. They evaluate the enactability of protocols under different infrastructure assumptions, e.g., with or without FIFO delivery. \citeauthor{Ferrando+19:enactability} do not consider the problem of role binding. 
Other early work on formally specifying \citep{Aamas-Enactment-03} and enacting \citep{Atal97} multiagent interactions also doesn't address role binding as an explicit concern.

Pippi's protocols can be enacted with the minimal set of infrastructure assumptions, namely that only sent messages be delivered. Further, Pippi's protocols can be interleaved without requiring to be composed (this is a property of information protocols \citep{JAIR-20:Langeval}), which is not a possibility with trace-based approaches. Role binding for a protocol can be interleaved with the enactment of the protocol, as shown in the \pname{Proposal} metaprotocol in Listing~\ref{prot:proposal}.

\citet{Chocron+Schorlemmer-20:vocabulary} consider the problem in open MAS where agents know they are participating in the same interaction (specified in temporal logic) but have different message vocabularies. They propose a method by which agents can dynamically align their vocabularies, that is, learn the mappings between their vocabularies. Pippi currently assumes a shared vocabulary (whatever is in the shared information protocols) but would benefit from alignment techniques in scenarios where the assumption does not hold. Arguably, aligning the ``same'' information protocol with different vocabularies is a more natural and challenging problem to address since information protocols can be more flexible than message ordering constraints specified in temporal logic.

\citet{Rocha+Brandao-19:IoT} apply multiagent systems to model dynamism in Internet of Things applications. They model devices via agents that enter and leave the system. Pippi enables realizing such dynamism in a general way with clean, modular representations. 

JADE \citep{Bergenti+20:JADE} supports the discovery of agents via a directory facilitator that provides a yellow pages service. JADE supports FIPA interaction protocols by providing endpoint implementations (classes and methods) that agents can use to implement interactions with others. However, JADE is not equipped to enable building multiagent systems off protocol specifications as Pippi's adapter and programming model enable. Further, discovery, although important, differs from role binding. In the wedding scenario, an agent can discover other agents looking for prospective partners via a directory service but would need to \emph{bind} one of them to Proposee to enact Proposal.  \citet{Briola+19:P2P-agent} demonstrate how JADE can be used to discover agents and implement a peer-to-peer system. As the foregoing examples demonstrate, Pippi supports both discovery and role binding and thus enables the realization of arbitrary applications as a peer-peer system.

\citepos{CarrieroGelernter92} tuple space approach for coordinating processes has been influential in multiagent systems. It features as the underlying coordination mechanism in CArtAgO \citep{Ricci+09:CArtAgO}, which itself is part of JaCaMo \citep{Boissier+13:JaCaMo}, which \citet{Baldoni+19:artifacts} use to support the implementation of commitment-based business processes. Tuple spaces (like logic programming) are attractive for their information-based abstractions (one works with tuples of information). However, they represent a shared-memory approach, which is not suitable for building decentralized MAS. It is widely argued that a tuple space decouples the readers and writers of information, but that argument holds only where the writers don't care who reads the information (as is the case in the classic ``readers and writers'' synchronization problem). In MAS, agents communicate with particular agents, e.g., the Proposer wants to communicate different information to the Proposee and to the Judge. Such coupling between agents is specified in the protocol. It would, however, be interesting to investigate if tuple spaces can be used to implement an agent's local state since it provides information-based abstractions.


\section*{Acknowledgments}
Thanks to the anonymous reviewers for their helpful comments.
Christie and Chopra were supported by EPSRC grant EP/N027965/1 (Turtles).
Singh was partially supported by the National Science Foundation under grant IIS-1908374.

\balance

\bibliographystyle{ACM-Reference-Format} 
\bibliography{Munindar,Amit,Samuel}

\end{document}